\newcommand{\sla}[1]{#1\!\!\!/}
\newcommand{\ba}{\begin{eqnarray}}
\newcommand{\ea}{\end{eqnarray}}

\documentclass[aps,preprint,showpacs,showkeys]{revtex4}
\usepackage{graphicx}
\usepackage{amssymb}

\begin{document}

\title{Effective Potential for Uniform Magnetic
Fields through Pauli Interaction}

\author{Hyun Kyu \surname{Lee}\footnote{hyunkyulee@hanyang.ac.kr} and Yongsung \surname{Yoon}\footnote{cem@hanyang.ac.kr}}

\affiliation{Department of Physics, Hanyang University, Seoul 133-791, Korea}

\begin{abstract}

We have calculated the explicit form of the real and imaginary parts of the effective potential for uniform magnetic
fields which interact with spin-1/2 fermions through the Pauli interaction. It is found that the non-vanishing
imaginary part develops for a magnetic field stronger than a critical field, whose strength is the ratio of the fermion
mass to its magnetic moment. This implies the instability of the uniform magnetic field beyond the critical field
strength to produce  fermion pairs with the production  rate density $w(x)=\frac{m^{4}}{24\pi}(\frac{|\mu
B|}{m}-1)^{3}(\frac{|\mu B|}{m}+3)$ in the presence of Pauli interaction.
\end{abstract}



\maketitle

\section{Introduction}

The interaction of charged spin-1/2 fermions with electromagnetic fields is described by the minimal coupling in the
form of Dirac equation. One of the interesting phenomena with strong electromagnetic fields is particle production. A
well known example is the Schwinger process, in which minimally interacting charged particles are created in pairs in
strong electric fields \cite{schwinger}. In a purely magnetic field configuration, however, it has been shown that the
production of minimally interacting fermion is not possible even with a spatial inhomogeneity \cite{dunn}. Therefore,
the pair production of minimally interacting particles is considered to be a purely electric effect.

Pauli introduced a non-minimal coupling of spin-1/2 particles with electromagnetic fields, which can be interpreted as
an effective interaction of fermions with an anomalous magnetic moment \cite{pauli,lavrov,ho}.  For the neutral
fermions with non-vanishing magnetic moments, it is the Pauli interaction through which  the electromagnetic
interaction can be probed. It is interesting to note that the inhomogeneity of the magnetic field, which couples
directly to the magnetic dipole moment through the Pauli interaction, plays a similar role analogous to the electric
field for a charged particles with the minimal coupling. The possibility of production of the neutral fermions in a
purely magnetic field configuration  with spatial inhomogeneity has been demonstrated in 2+1 dimension \cite{lin}, and
recently the production rate in 3+1 dimension  has been calculated explicitly for the magnetic fields with a spatial
inhomogeneity of a critical value \cite{leeyoon}.

The purpose of this paper is to discuss further the possibility of
the fermion production under a uniform magnetic field when it
becomes stronger than the critical field whose strength is the ratio
of the fermion mass to its magnetic moment.  We consider a neutral
fermion but with a magnetic moment $\mu$ with Pauli interaction. The
energy eigenvalues of the fermion is given by
\begin{equation}
E=\pm \sqrt{p_{l}^{2}+(\sqrt{m^{2}+p_{t}^{2}} \pm \mu B)^{2}},
\label{energy}
\end{equation}
where $p_{l}$ and $p_{t}$ are respectively the longitudinal and the
transversal momentum to the magnetic field direction. One can see
that, for a  critical magnetic field $B_{c}=\frac{m}{\mu}$, the
energy gap between the positive and the negative energy states
disappears. This indicates the possible instability of magnetic
field configurations even in uniform magnetic fields.    We have
calculated the effective potential of uniform magnetic fields which
interact with spin-1/2 fermions through the Pauli interaction. For a
magnetic field weaker than the critical field, the effective
potential is real as expected. However, for a magnetic field
stronger than the critical field, it is found that the imaginary
part of the effective potential does not vanish.  This implies that
a uniform magnetic field becomes unstable to produce the fermion
pairs in vacuum when it is stronger than the critical field. It
should be noted  that the pair production in uniform magnetic fields
is not due to the  tunneling process as in Schwinger process
overcoming the energy gap, $2m$, but due to the disappearance of the
energy gap in Eq.(\ref{energy}) for the critical field strength. The
difference is also manifested in different  functional forms of the
pair production rates. It is found that the production rate  takes a
quartic form which is quite different from the exponential form of
Schwinger process.

The calculation of the effective potential for uniform magnetic
fields induced by a neutral fermion, which is assumed to be
interacting with background electromagnetic field through Pauli
coupling,  is discussed in Section II and the results are summarized
in Section III.

\section{Effective Potential for Uniform Magnetic Fields Induced by Fermions with Pauli Interaction}

The Dirac Lagrangian of a neutral fermion with Pauli interaction is given by
\begin{equation}
{\cal L} =
\bar{\psi}(\sla{p}+\frac{\mu}{2}\sigma^{\mu\nu}F_{\mu\nu}-m)\psi,\label{pauli}
\end{equation}
where $\sigma^{\mu\nu}=\frac{i}{2}[\gamma^{\mu},\gamma^{\nu}], ~~g_{\mu\nu}=(+,-,-,-)$.  $\mu$ in the Pauli term
measures the magnitude of the magnetic moment of the neutral fermion. The corresponding Hamiltonian is given by
\begin{equation}
H = \vec{\alpha}\cdot (\vec{p}-i\mu\beta\vec{E})+\beta(m-\mu\vec{\sigma}\cdot\vec{B}), \label{hamiltonian}
\end{equation}
where $ \sigma^{i}=\frac{1}{2}\epsilon^{ijk}\sigma^{jk}$. The energy
eigenvalues of the Hamiltonian Eq.($\ref{hamiltonian}$) are given by
Eq.(\ref{energy}). One can see that, for a magnetic field stronger
than the critical field $B_{c}=\frac{m}{\mu}$, the energy gap
between the positive and the negative energy states disappears. This
indicates the possible instability of magnetic field configuration.
On the other hand, the energy eigenvalues of minimally interacting
charged fermions without an anomalous magnetic moment are
\begin{equation}
E=\pm \sqrt{p_{l}^{2}+m^{2}+|e|B(2n+1-{\rm sgn}(e)\hat{s})}, \label{energy2}
\end{equation}
where $n=0,1,2,\ldots$, and $\hat{s}=\pm 1$ are spin projections along the magnetic field \cite{khalilov}. It should be
pointed out that  zero energy states do not exist even for a strong magnetic field and no particle production of
minimally interacting fermions in pure magnetic fields can be attributed to this finite energy gap \cite{gap}.

The effective potential $ V_{\rm eff}(A)$ for a background electromagnetic vector potential $A_{\mu}$ can be
obtained by integrating out the fermions:
\begin{eqnarray}
-i \int d^{4}x V_{\rm eff}(A[x]) =\int d^{4}x <x|tr \ln
[(\sla{p}+\frac{\mu}{2}\sigma^{\mu\nu}F_{\mu\nu}-m)\frac{1}{\sla{p}-m} ]|x>, \label{veff}
\end{eqnarray}
where $F_{\mu\nu}=\partial_{\mu}A_{\nu} - \partial_{\nu}A_{\mu}$, and $tr$ denotes the trace over Dirac algebra. The
decay probability of the background magnetic field into the neutral fermions is related to the imaginary part of the
effective potential $V_{\rm eff}(A)$,
\begin{equation}
P = 1-|e^{i \int d^{4}x  V_{\rm eff}(A[x])}|^{2} = 1-e^{-2\Im \int d^{3}x dt V_{\rm eff}(A[x])}.
\end{equation}
That is, the twice of the imaginary part of the effective potential $V_{\rm eff}(A[x])$ is the fermion production rate
per unit volume \cite{field}: $w(x)= 2\Im (V_{\rm eff}(A[x]))$ for small probabilities.

For a uniform magnetic field configuration such that $\vec{B}=B \hat{z}$, the integral form of the effective potential
Eq.($\ref{veff}$) is obtained as \cite{leeyoon}
\begin{equation}
V_{\rm eff}=-\frac{(\mu B)^{2}}{4\pi^{2}} \int^{\infty}_{0}\frac{ds}{s^{2}} [ i\int^{1}_{0}d\xi(1-\xi)e^{i(\mu B)^{2}
\xi^{2}s}-\frac{i}{2}+\frac{(\mu B)^{2}s}{12} ] e^{-im^{2}s}. \label{VC}
\end{equation}
The integration Eq.($\ref{VC}$) can be done explicitly. Introducing dimensionless parameters, $t=m^{2}s$ and
$\beta=\frac{|\mu B|}{m}$, the imaginary part of the effective potential Eq.($\ref{VC}$) can be  written as
\begin{eqnarray}
\Im(V_{\rm eff})&=&-\frac{m^{4}\beta^{2}}{4\pi^{2}} \int^{1}_{0}d\xi(1-\xi)\int^{\infty}_{0}\frac{dt}{t^{2}}[
\cos(\beta^{2} \xi^{2}-1)t-\cos(t)-\beta^{2}\xi^{2}t \sin(t)] \nonumber \\
 ~&=& -\frac{m^{4}\beta^{2}}{8\pi} \int^{1}_{0}d\xi(1-\xi)[1 - \beta^{2}\xi^{2}
-|1-\beta^{2}\xi^{2}|  ]. \label{im}
\end{eqnarray}

For a magnetic field weaker than the critical field, $\beta \leq 1$, the integration Eq.($\ref{im}$) vanishes. It can
be also verified by a contour integration. For the  magnetic fields weaker than the critical field $B_c=m/\mu$, using a
contour integration in the fourth quadrant, the integration can be done along the negative imaginary axis giving the
finite real effective action as
\begin{equation}
V_{\rm eff}=-\frac{(\mu B)^{2}}{4\pi^{2}} \int^{\infty}_{0}\frac{ds}{s^{2}}[\frac{1}{2}+\frac{(\mu
B)^{2}s}{12}-\int^{1}_{0}d\xi(1-\xi)e^{(\mu B)^{2} \xi^{2}s} ] e^{-m^{2}s}. \label{VCR}
\end{equation}
Therefore, one can see that the uniform magnetic fields weaker than the critical field are stable as expected.

However, for a magnetic field stronger than the critical field, $\beta > 1$, the imaginary part of the effective
potential does not vanish, but takes a quartic form:
\begin{equation}
\Im(V_{\rm eff})=\frac{1}{48\pi}(|\mu B|-m)^{3}(|\mu B|+3m), \label{im2}
\end{equation}
which is the half of the particle production rate density $w$ for a magnetic field stronger than the critical field
$B_c=m/\mu$. Therefore, the uniform magnetic fields stronger than the critical field are unstable, and reduce the field
strengths producing the fermion pairs due to the disappearance of the energy gap.

The real part of the effective potential Eq.($\ref{VC}$) can be calculated explicitly as well,
\begin{eqnarray}
\Re(V_{\rm eff})&=&-\frac{m^{4}\beta^{2}}{4\pi^{2}} \int^{1}_{0}d\xi(1-\xi)\int^{\infty}_{0}\frac{dt}{t^{2}} [
\sin(1-\beta^{2}\xi^{2})t -\sin(t) +\beta^{2}\xi^{2}t\cos(t) ]  \label{re} \\ ~&=& \left\{ \begin{array}{ll}
\frac{m^{4}}{288\pi^{2}} [13\beta^{4}-78\beta^{2}+96\beta\tanh^{-1}(\beta)-6(\beta^{4}-6\beta^{2}-3)\ln(1-\beta^{2})],
& {\rm for}~ \beta \leq 1
\\ \frac{m^{4}}{288\pi^{2}}
[13\beta^{4}-78\beta^{2}+96\beta\coth^{-1}(\beta)-6(\beta^{4}-6\beta^{2}-3)\ln(\beta^{2}-1)], & {\rm for}~ \beta \geq
1.
\end{array} \right. \nonumber
\end{eqnarray}
For a weak field, $\beta \ll 1$, Eq.($\ref{re}$) approximates to $\frac{(\mu B)^{6}}{240\pi^{2}m^{2}}$, and for the
critical field, $\beta=1$, $\Re(V_{\rm eff})=(96\ln2-65)\frac{m^{4}}{288\pi^{2}}$. The real and imaginary parts of the
effective potential with respect to the magnetic field strength are shown in FIG.1 in the unit of
$\frac{m^{4}}{48\pi^2}$.

\begin{figure}[htp]
\begin{center}
\includegraphics[height=2.5in,width=3.0in]{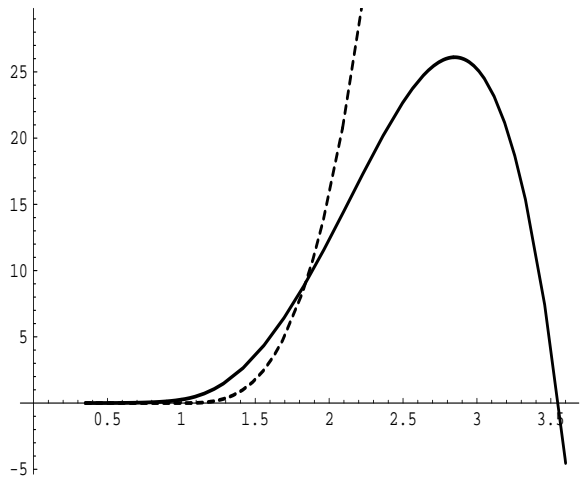}
\caption{Effective potential of the uniform magnetic field B induced by neutral fermions with a magnetic moment:
vertical axis is $V_{\rm eff}$ in the unit of $\frac{m^{4}}{48\pi^2}$ (the dashed line is for the imaginary part and
the solid line is for the real part), horizontal axis is $\beta(=|\mu B|/m$).}
\end{center}
\end{figure}

So far, we have considered only the neutral fermions with a magnetic dipole moment. It is also interesting to see how
the instability due to the Pauli interaction is  affected when the minimal coupling is turned on in addition to the
Pauli interaction. Let us consider an effective Lagrangian, which might describe a fermion endowed with  a
non-vanishing electric charge $e$ and as well as with a magnetic dipole moment $\mu$, given by
\begin{equation}
{\cal L} = \bar{\psi}(\sla{p}-e\sla{A}+\frac{\mu}{2}\sigma^{\mu\nu}F_{\mu\nu}-m)\psi.
\label{pauli_e}
\end{equation}
In QED,  $\mu$ could be identified as the Schwinger's anomalous magnetic moment $\mu_a =
\frac{\alpha}{2\pi}\frac{e}{2m} $,  which comes from the 1-loop radiative corrections \cite{field}. However the full
calculation of the QED radiative correction for strong magnetic fields \cite{radiative} shows that  the Pauli term
description of the Schwinger's anomalous magnetic moment is valid only for weak magnetic fields such that $B \ll
m^{2}/e$, which is much smaller than the critical field for the anomalous magnetic moment defined by $B_c= m/\mu_a$.

In this work, however, we consider a  model in which the electric charge $e$ and the magnetic dipole moment $\mu$ are
two independent couplings such that  the Pauli term description of the magnetic moment  is valid up to the critical
magnetic field. Physically it corresponds to the  condition that  the radiative corrections due to the minimal coupling
does not dominate over the Pauli term up to the critical magnetic field.  Then, the effective potential for a fermion
described by Eq.(\ref{pauli_e}) is calculated as \cite{leeyoon2}
\begin{eqnarray}
V_{\rm eff}= &-& \frac{1}{8\pi^2}\int^{\infty}_{0}\frac{ds}{s^{2}}[ |eB|\coth(|eB|s)-\frac{1}{s}-\frac{(eB)^{2}s}{3}
]e^{-m^{2}s} \label{VCC} \\ ~&-& \frac{(\mu B)^{2}}{4\pi^{2}} \int^{\infty}_{0}\frac{ds}{s^{2}}[ i|eB|s \cot(|eB|s)
\int^{1}_{0}d\xi(1-\xi)e^{i(\mu B)^{2} \xi^{2}s}-\frac{i}{2}+\frac{(\mu B)^{2}s}{12} ] e^{-im^{2}s}, \nonumber
\end{eqnarray}
where the first integral is the effective potential for a minimally interacting charged fermion in a uniform magnetic
field \cite{dunn}, and the second integral is the contribution from the magnetic moment $\mu$.

For a magnetic field weaker than the critical field $B_{c}=m/\mu$, the effective potential Eq.($\ref{VCC}$) is real as
expected because the $s$ integration of the second integral can be done along the negative imaginary axis in the fourth
quadrant,
\begin{eqnarray}
V_{\rm eff}=&-&\frac{1}{8\pi^2}\int^{\infty}_{0}\frac{ds}{s^{2}}[ |eB|\coth(|eB|s)-\frac{1}{s}-\frac{(eB)^{2}s}{3}
]e^{-m^{2}s} \label{VRE} \\ ~&-&\frac{(\mu B)^{2}}{4\pi^{2}} \int^{\infty}_{0}\frac{ds}{s^{2}}[ \frac{1}{2}+\frac{(\mu
B)^{2}s}{12}- |eB|\coth(|eB|s) \int^{1}_{0}d\xi(1-\xi)e^{(\mu B)^{2} \xi^{2}s} ] e^{-m^{2}s}. \nonumber
\end{eqnarray}

For a magnetic field stronger than the critical field, isolating singularities at $s=0$ in the
second integral of Eq.($\ref{VCC}$), we can rewrite the effective potential as
\begin{eqnarray}
V_{\rm eff}=&-&\frac{1}{8\pi^2}\int^{\infty}_{0}\frac{ds}{s^{2}}[ |eB|\coth(|eB|s)-\frac{1}{s}-\frac{(eB)^{2}s}{3}
 ] e^{-m^{2}s} \nonumber \\ ~&-&\frac{(\mu B)^{2}}{4\pi^{2}} \int^{\infty}_{0}\frac{ds}{s^{2}}[ i
\int^{1}_{0}d\xi(1-\xi)e^{i(\mu B)^{2} \xi^{2}s}-\frac{i}{2}+\frac{(\mu B)^{2}s}{12} ] e^{-im^{2}s} \nonumber
\\ ~&+& \frac{(\mu B)^{2}}{4\pi^{2}}
\int^{\infty}_{0}ds \{|eB|s \coth(|eB|s) -1 \} \nonumber \\ ~&~&~[\int^{\xi_{0}}_{0}d\xi(1-\xi)e^{-(m^{2}-(\mu
B)^{2}\xi^{2})s} - \int^{1}_{\xi_{0}}d\xi(1-\xi)e^{-((\mu B)^{2}\xi^{2}-m^{2})s}], \label{VCES}
\end{eqnarray}
and $\xi_{0} \equiv |\frac{m}{\mu B}| < 1$. In Eq.($\ref{VCES}$), the first integral is known to be real, and it is
straightforward to verify that the third integral is real.  The second integral in Eq.($\ref{VCES}$) is exactly the
effective potential Eq.($\ref{VC}$) for neutral fermions. Thus, the imaginary part of the effective potential comes
only from the contributions of the magnetic moment, and the production rate density Eq.($\ref{im2}$) calculated for
neutral fermions is also valid for charged fermions with the magnetic moment $\mu$.  Hence there is no effect of the
electric charge with the minimal coupling on the instability due to the magnetic moment through the Pauli interaction
for a fermion described by Eq.(\ref{pauli_e}).

\section{Discussion}
For charged  fermions, which couple to the electromagnetic field through minimal coupling, it has been well known that
pair creation is not possible in purely magnetic field configurations \cite{schwinger,dunn,gap}.  In this work, we
discuss  the possibility that particles can be created in a strong enough magnetic field as a purely magnetic effect.
Introducing the magnetic moment of neutral spin-1/2 fermions through Pauli interaction, it has been shown that
production of neutral fermions is possible in a purely magnetic field configuration provided that the gradient of the
magnetic field is extremely strong \cite{lin,leeyoon}. However, the particle production in uniform magnetic fields has
not yet been addressed properly.  We calculate explicitly the real and imaginary part of the effective potential for a
uniform magnetic field, by integrating out the fermions with a magnetic moment which couples to the magnetic fields
through Pauli interaction. We have shown explicitly that the imaginary part of the effective potential develops when
the uniform magnetic fields are stronger than the critical field $B_c=\frac{m}{\mu}$. Hence the magnetic field
background stronger than $B_c$ is unstable to produce the fermion pairs.  We have calculated the production rate
density $w$ of the fermions as $w=\frac{m^{4}}{24\pi}(\frac{|\mu B|}{m}-1)^{3}(\frac{|\mu B|}{m}+3)$.  One can  note
that this result is quite different from the exponential form of Schwinger process.  The main reason for this
difference is that the pair production in uniform magnetic fields is not due to the tunneling process as in Schwinger
process overcoming the energy gap, $2m$, but due to the disappearance of the energy gap.

One of the immediate application of the particle production
mechanism discussed in this work might be a particle creation in the
vicinity of the compact objects in the strong explosive
astrophysical phenomena, where the extraordinarily strong magnetic
fields ($> 10^{15}$G) are expected and the environment is considered
to be magnetically dominant.  Of course it depends on whether  there
is any physical model in which the  neutral fermion considered  is
described by Pauli interaction up to the critical magnetic fields.

This work was supported by grant No. (R01-2006-000-10651-0) from the Basic Research Program of the Korea Science \&
Engineering Foundation.

\end{document}